\documentclass[a4paper,11pt]{article} 

\pdfoutput=1

\usepackage{jheppub}
\usepackage[T1]{fontenc} 
\usepackage{accents}

\title{\boldmath  Effect of cross-section models on the validity of sterile neutrino mixing limits}

\author{Patrick Stowell,} 
\author{Callum Wilkinson,}
\author{Susan Cartwright.} 
\affiliation{Department of Physics and Astronomy, University of Sheffield, \\
		Hicks Building, Hounsfield Road, Sheffield, S3 7RH, United Kingdom}

\emailAdd{p.stowell@sheffield.ac.uk}

\abstract{ 
Charged-Current Quasi-Elastic (CCQE) neutrino scattering is the signal channel for sterile neutrino
oscillation experiments. Recent cross-section measurements have made it clear that the current
understanding of this channel in the few-GeV region is incomplete, and several sophisticated
theoretical models have been proposed to tackle this issue, although it is not clear which model
best describes the global dataset. In this paper we argue that the current uncertainty surrounding
CCQE cross-sections is a serious problem for experiments seeking to produce sterile neutrino limits.
We perform a sterile neutrino analysis with published MINER$\nu$A data as an illustrative example.
We highlight the need for caution in interpreting sterile neutrino limits given the context of
incomplete cross-section model information.
}

\notoc 

\begin{document} 
\maketitle 
\flushbottom

\section{Introduction} 
\label{sec:intro}

Accelerator neutrino experiments in the few-GeV region, with detectors at short baselines, are used
both to constrain sterile neutrino mixing models and to measure neutrino-nucleus scattering
cross-sections. As the measured quantities are event rates -- the flux multiplied by the
cross-section -- the measurement of either relies on some assumption about the other.

For a long time, relativistic Fermi gas (RFG) models \cite{smith-moniz72} have been used to describe
charged-current neutrino-nucleus scattering in generators \cite{zeller12}. The only free parameter
unconstrained by electron scattering data in these models is the axial mass, $M_\mathrm{A}$, which
was well-constrained to be $M_\mathrm{A}=1.014\pm0.014\mbox{ GeV}$ \cite{bodek08vec} from deuterium
scattering \cite{bodek08deut} and pion electro-production data \cite{liesenfield99pion}. It was
therefore believed that the Charged-Current Quasi-Elastic (CCQE) cross-section was well-understood;
however, recent neutrino-nucleus scattering data on heavy nuclear targets have produced much higher
cross-sections, and much higher axial-mass values in this simple cross-section parametrisation
\cite{minib10neut,minib13anti,k2k06neut,adamson2014pgc,t2k14ccqe}. This discrepancy is thought to
result from additional nuclear effects which are not included in the RFG models \cite{sobczyk10}.
This has led in recent years to the development of more sophisticated models to explain the
incompatibility between datasets. These models differ significantly in their prediction of outgoing
particle kinematic distributions, and as such, the state of neutrino-nucleus scattering
cross-sections in the few-GeV region cannot be said to be well understood.

In this analysis we investigate the effect that different cross-section models of the CCQE
interaction channel have on the limits produced by a short-baseline muon-neutrino disappearance
analysis using a 3+1 mixing model. The cross-section models investigated are a small range of those
currently available in generators. The MINER$\nu$A CCQE cross-section data in neutrino and
antineutrino modes \cite{fields13,fiorentini13} is used as an illustrative example, though the
conclusions of this work apply to any sterile neutrino measurements made with accelerator neutrino
beams in the few-GeV region. We show that the choice of cross-section model has a significant impact
on the sterile neutrino confidence limits produced, and argue that the current uncertainty on the
CCQE cross-section makes sterile neutrino limits in this energy range difficult to interpret.
This work builds on work done in \cite{wilkinson14, wilkinson14pca} to show that modifications to
$M_\mathrm{A}$ in the RFG model can affect the neutrino limits produced by sterile analyses. It
complements other work investigating the effect that uncertainties in the cross-section models have
on the reconstructed energy \cite{ohmar13effects}, fitted limits on $\delta_{CP}$
\cite{meloni12t2kfit}, and atmospheric mixing limits \cite{coloma14} in a three neutrino framework.

Whilst a fake data study would have been equally valid for the purpose of this analysis, we chose to
use public MINER$\nu$A CCQE cross-section data, as a sterile neutrino fit of this kind has not yet
been performed on these datasets. The NuWro Monte Carlo event generator \cite{nuwro09} was used to
produce differential cross-sections from initial event rate predictions for the CCQE cross-section
models detailed in Section \ref{sec:cross}. Sterile neutrino induced biases to these predictions
were produced by folding in a muon-neutrino survival probability under the 3+1 mixing model
described in Section \ref{sec:sterile}. Each sterile hypothesis was then fitted to the MINER$\nu$A
dataset detailed in Section \ref{sec:minerva} and a $\chi^2$ statistic was calculated as described
in Section \ref{sec:method}. For each of the cross-section models we investigated, $\chi^2$ scans
were performed in the $\sin^2 2\theta_{\mu\mu}-\Delta m^{2}_{24}$ plane. The resulting confidence
intervals are discussed in Section \ref{sec:conclusion}.

\section{Cross-section models} 
\label{sec:cross}

This section describes the key features of the cross-section models considered in this analysis.
There are three nuclear models, described in Section \ref{sec:nucmodels}, one of which is the
familiar Smith-Moniz RFG model \cite{smith-moniz72} used in many generators and past analyses.
Models of additional nuclear effects are described in Section \ref{sec:additionaleffects}.

\subsection{Underlying nuclear model} 
\label{sec:nucmodels}
Dipole axial form factors \cite{multinucleon} and BBBA05 modifications \cite{bbba05} to vector form
factors were used consistently for all of the models described in this section.
\\
\textbf{Relativistic Fermi Gas (RFG):}
Nucleons are treated as quasi-free with a nucleus-dependent Fermi momentum and constant binding
energy, $E_b$ \cite{smith-moniz72}. This model uses the impulse approximation where the neutrino
interacts with one nucleon only. In the RFG model all states up to the Fermi momentum are filled, so
interactions where the outgoing nucleon is not outside the RFG distribution are Pauli blocked. The
Bodek-Ritchie modification to the RFG model is included, which adds a higher momentum contribution
due to short-range correlations between nucleons \cite{bodekritchie81fermi}.
\\
\textbf{Benhar Spectral Function (SF):} 
A nucleus-dependent description of nucleon kinematics within the nucleus, in terms of its removal
energy and momentum \cite{benhar10spectral}. Approximately 20\% of the cross-section is due to
short-range correlations of nucleons (quasi-deuterons). The impulse approximation is used
consistently; the interaction is with a single nucleon even for correlated states. Pauli blocking is
approximated by a nucleon-dependent cut-off \cite{benhar94spectral}. 
\\
\textbf{Local Fermi Gas (LFG):} 
Similar to the RFG but the binding energy, $E_b$, varies with the nucleon position within the
nucleus, producing a more realistic Pauli blocking effect \cite{leitner09LFG,leitner06LFG}.
\\
\subsection{Nuclear effects}
\label{sec:additionaleffects}
Recent models attempt to explain the large MiniBooNE axial mass value in terms of modifications to
CCQE interactions within the nucleus. We consider three such enhancements to the standard model.
\\
\textbf{Transverse Enhancement Model (TEM):}
A four-momentum transfer dependent modification to the magnetic form factor \cite{bodek11TEM}. The
modification is obtained by fitting to an experimentally observed excess in the ratio of transverse
to longitudinal quasi-elastic response functions from electron scattering data
\cite{donnelly99scaling}.
\\
\textbf{Random Phase Approximation (RPA):}
A modification to the quasi-elastic propagator, which accounts for long range nucleon-nucleon
correlations within the nuclear medium \cite{nieves04RPA}.
\\
\textbf{Nieves multi-nucleon interaction model:}
A microscopic model that sums over possible $W$ boson absorption modes, where the interaction is
with two or three nucleons \cite{nieves06model}. Note that this is an explicit contribution beyond
the impulse approximation which has final states that are largely indistinguishable from CCQE
interactions (CCQE-like), and will therefore enhance reported CCQE cross-section measurements.
\subsection{The Nieves model} The Nieves model \cite{nieves06model} is a consistent description of
the CCQE-like cross-section which incorporates the LFG, the RPA, and the Nieves multi-nucleon
interaction model. From now on the ``Nieves model'' will refer to this combination.

\section{Sterile models}
\label{sec:sterile}

3+1 neutrino models extend the $3\times3$ PMNS matrix by including an additional, predominantly
sterile, mass state which is heavier than the other three neutrinos. Over short baselines, the three
active mass states can be approximated as degenerate, and a two neutrino mixing equation can be used
to describe mixing between any of the three active states and the larger sterile mass state. The
survival probability of a muon (anti-)neutrino can then be calculated using \cite{conrad12short}.
\begin{align}
	 \label{eq:shortbaseline}  P\left( \accentset{(-)}{\nu}_{\mu} \rightarrow  \accentset{(-)}\nu_{\mu} \right) &=
	1 - \sin^{2}2\theta_{\mu \mu} 
	\sin^{2}\left(\frac{1.265~\Delta m_{24}^2 \mbox{[eV$^2$]}~ L\mbox{[km]}}{E_\nu \mbox{[GeV]}}\right) 
	\,, 
\end{align}
where
\begin{align}
	 \sin^2 2\theta_{\mu \mu} = 4(1 - |U_{\mu 4}|^2)|U_{\mu 4}|^2 \,. 
\end{align}

\section{MINER$\nu$A CCQE data}
\label{sec:minerva}

This analysis uses the MINER$\nu$A $\nu_\mu$ and $\bar{\nu}_\mu$ CCQE cross-section measurements
\cite{fields13,fiorentini13}. The data were taken on a CH target and are presented as a differential
in reconstructed four-momentum transfer, $Q^2_{QE}$. The key experimental details are summarised in
Table \ref{tab:analys} \cite{minerva14spec}. The public data release includes the full covariance
matrix including correlations between the two datasets.

The reconstructed neutrino energy, $E_\nu^{QE}$, and four-momentum transfer, $Q^2_{QE}$, are derived
from the outgoing lepton kinematics ($E_\mu, p_\mu, \theta_\mu$) and the measured target binding
energy $E_b$ for the target by assuming the pure two-body kinematics of the RFG model:
\begin{align} 
	\label{eq:energy}
	 E_{\nu}^{QE} &= 
	\frac{m_n^2 - (m_p - E_b)^2 - m_\mu^2 + 2(m_p - E_b)E_\mu}{2(m_p - E_b - E_\mu + p_\mu \cos \theta_\mu)}
	 \,, \\ \quad \notag \\
	\label{eq:momentumtransfer} Q^2_{QE} &= 
	2E_\nu^{QE} (E_\mu - p_\mu \cos \theta_\mu) - m_\mu^2 \,.
\end{align}

\begin{table}[tbp] 
	\centering
	\begin{tabular}{ | l | c c | } 
		\hline 
		Neutrino Run & $\bar{\nu}_\mu$ & $\nu_\mu$ \\ 
		\hline
		Distance to target, $L$ (km) & 1.04 & 1.04 \\ 
		Energy range (GeV) & 1.5 $\leq E_\nu \leq 10.0$ & 1.5 $\leq E_\nu \leq 10.0$ \\
		Protons on target (POT) & $1.014 \times 10^{20}$ & $9.42 \times 10^{19}$ \\
		Integrated flux ($\nu$ cm$^{-2}$ POT$^{-1}$) & $2.429\times 10^{-8}$ & $2.916 \times 10^{-8}$ \\
		Target material & CH & CH \\
		Binding energy (MeV)  & 30 & 34  \\
		\hline 
	\end{tabular}
	\caption{Specifications of the MINER$\nu$A datasets used in this
		    analysis. The cross-correlations between the neutrino and antineutrino datasets provided in refs
		    \protect\cite{fields13,fiorentini13} allowed both datasets to be fitted simultaneously. The distance to target
		    was approximated as the distance from the NuMI target to the MINOS near detector \protect\cite{minervatech,minostech}.} 
	\label{tab:analys}
\end{table}

\section{Fitting method} 
\label{sec:method}

We used NuWro to make Monte-Carlo (MC) comparisons with the MINER$\nu$A datasets for each of the
cross-section models. Sterile neutrino induced biases were introduced by re-weighting the flux. This
approach allows large samples to be generated with minimal computational overhead. A $\chi^2$
minimization using the MINUIT \cite{minuit75} fitting package was used to determine best fit sterile
parameters and calculate limits in the sterile mixing plane.

MC events for each cross-section model were initially generated with a flat true neutrino energy
($E_\nu$) distribution across the experimental range.

The effect of a sterile neutrino is to modify the effective flux since sterile neutrinos do not
interact. We re-weight the effective MINER$\nu$A flux according to the survival probability
for a given sterile hypothesis and recalculate the derived cross-section according to steps
\ref{enum:method1}--\ref{enum:method2} shown below.

\begin{enumerate} 
	\item \label{enum:method1}Events were binned into a histogram $R(Q^2_{QE},E_\nu)$ where $Q^2_{QE}$ was calculated using Equation \eqref{eq:momentumtransfer}.
	\item $R(Q^2_{QE},E_\nu)$ was normalised to the total MC cross-section $\sigma^{MC}$.
	\item $R(Q^2_{QE},E_\nu)$ was weighted to the published MINER$\nu$A flux distribution $\Phi$.
	\item The survival probability for the $j^{th}$ $E_\nu$ bin,  $P(\Delta m_{24}^2,\theta_{\mu\mu},E_j)$, was calculated by averaging Equation \eqref{eq:shortbaseline} over 40 
	equally spaced points within the bin.
	\item $R(Q^2_{QE},E_\nu)$ was multiplied by $P\left(\Delta m_{24}^2,\theta_{\mu\mu},E_j\right)$ introducing a sterile bias.
	\item  \label{enum:method2}A cross-section histogram $B(Q^2_{QE})$ was created by projecting $R(Q^2_{QE},E_\nu)$ onto the $Q^2_{QE}$ axis. 
\end{enumerate} 
The $i^{th}$ $Q^2_{QE}$ bin of $B(Q^2_{QE})$ is thus given by
\begin{align}
	\label{eq:bincontents} 
	B_{i} = d\sigma_i^S = \sum_j \left( \frac{R_{ij}}{\sum_{kl} R_{kl}} \times \sigma^{MC}
	\times \frac{\Phi(E_j)}{\sum_k \Phi(E_k)} \times P_{j}\left(\Delta m_{24}^2,\theta_{\mu\mu}, E_j\right) \right) \,.
\end{align}

To reduce the statistical error from the MC to negligible levels, a large number of events ($10^7$)
were generated for each cross-section model. The statistical error on each $Q^2_{QE}$ bin is then
less than $0.1\%$.

\begin{figure}[tbp] 
	\centering
	\includegraphics[width=.495\textwidth,trim= 18 3 5 15,clip]{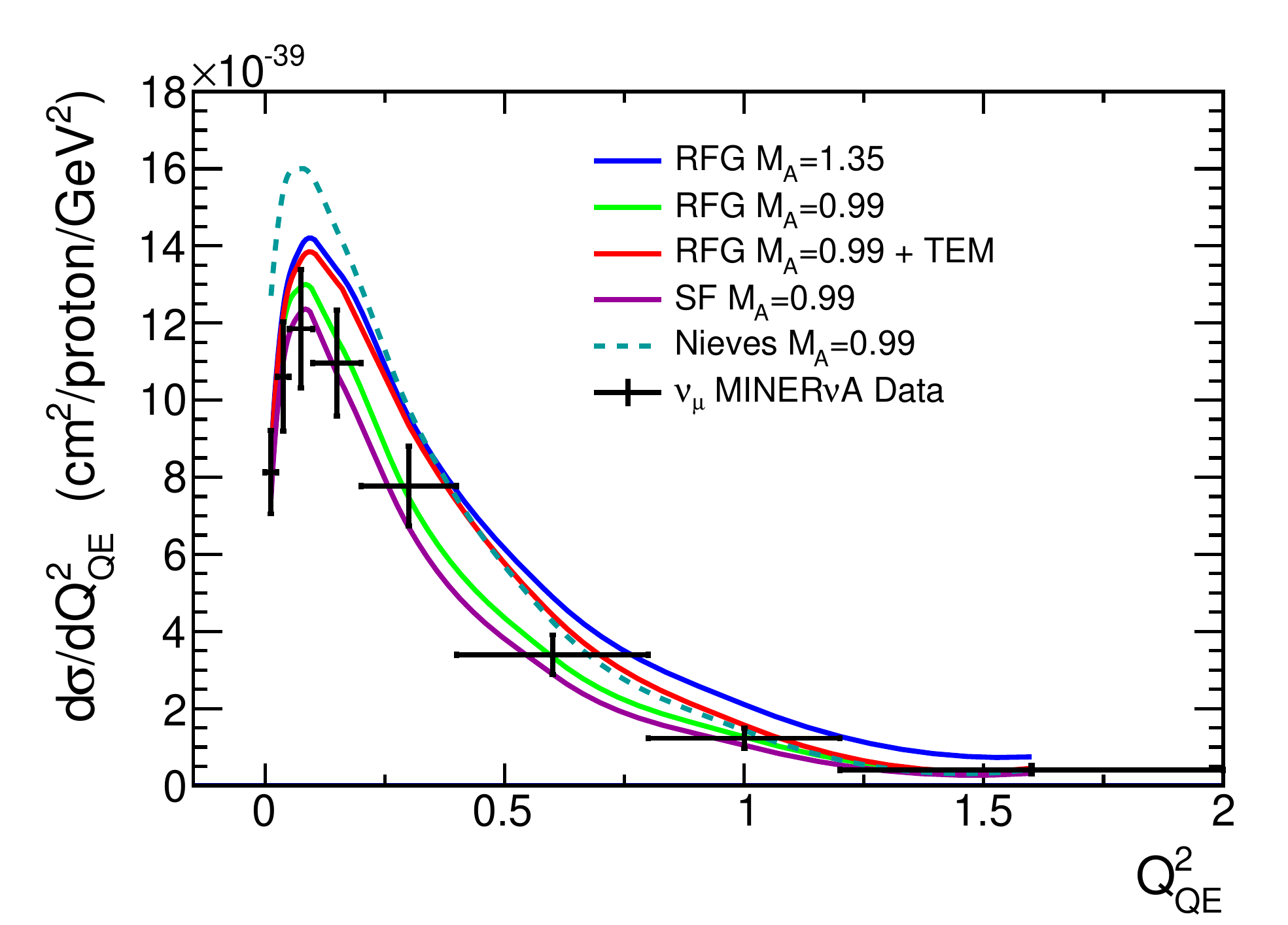} 
	\hfill
	\includegraphics[width=.495\textwidth,trim= 18 3 5 15,clip]{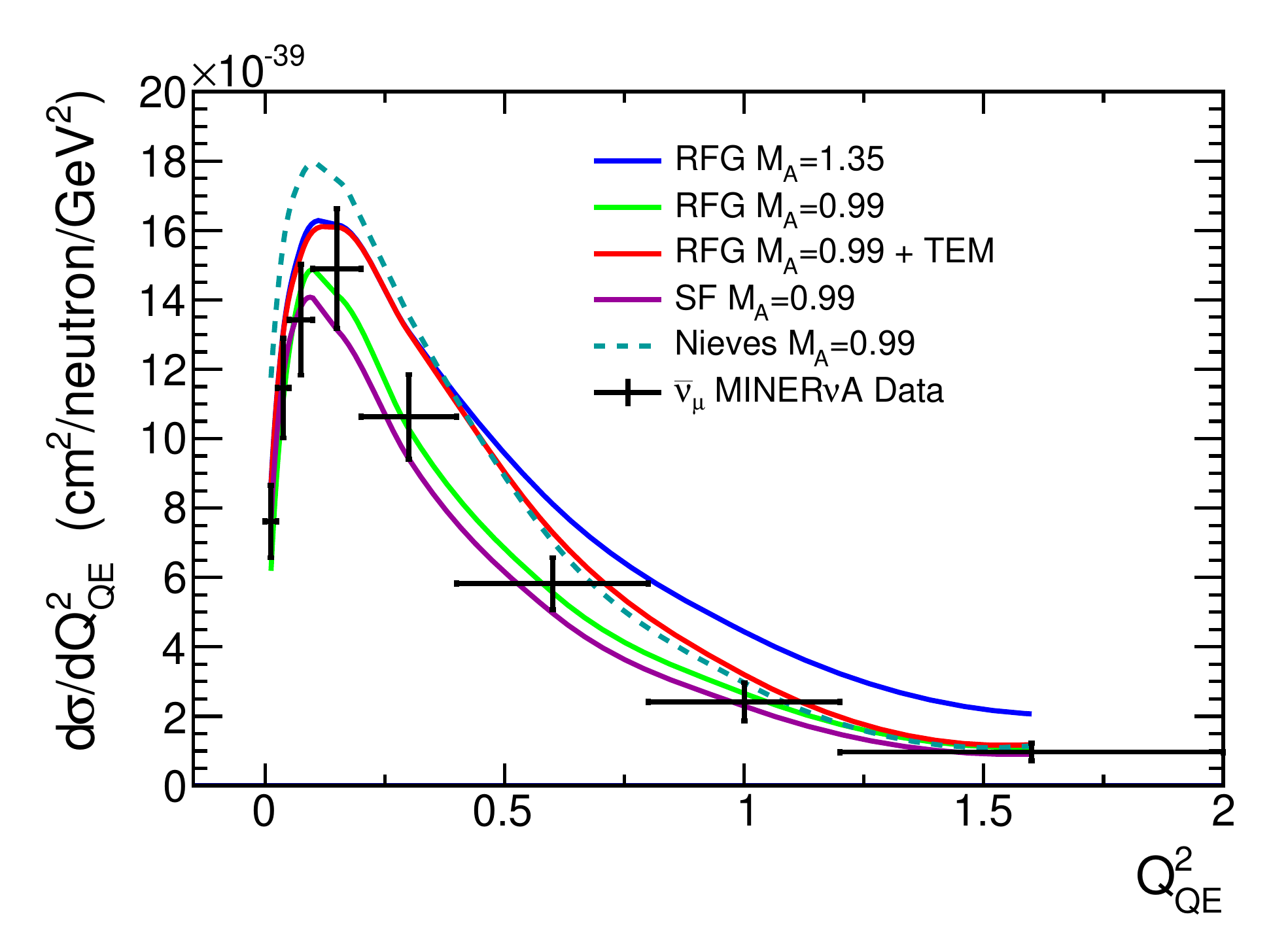} 
	
	\includegraphics[width=.49\textwidth, trim = 18 3 5 15,clip]{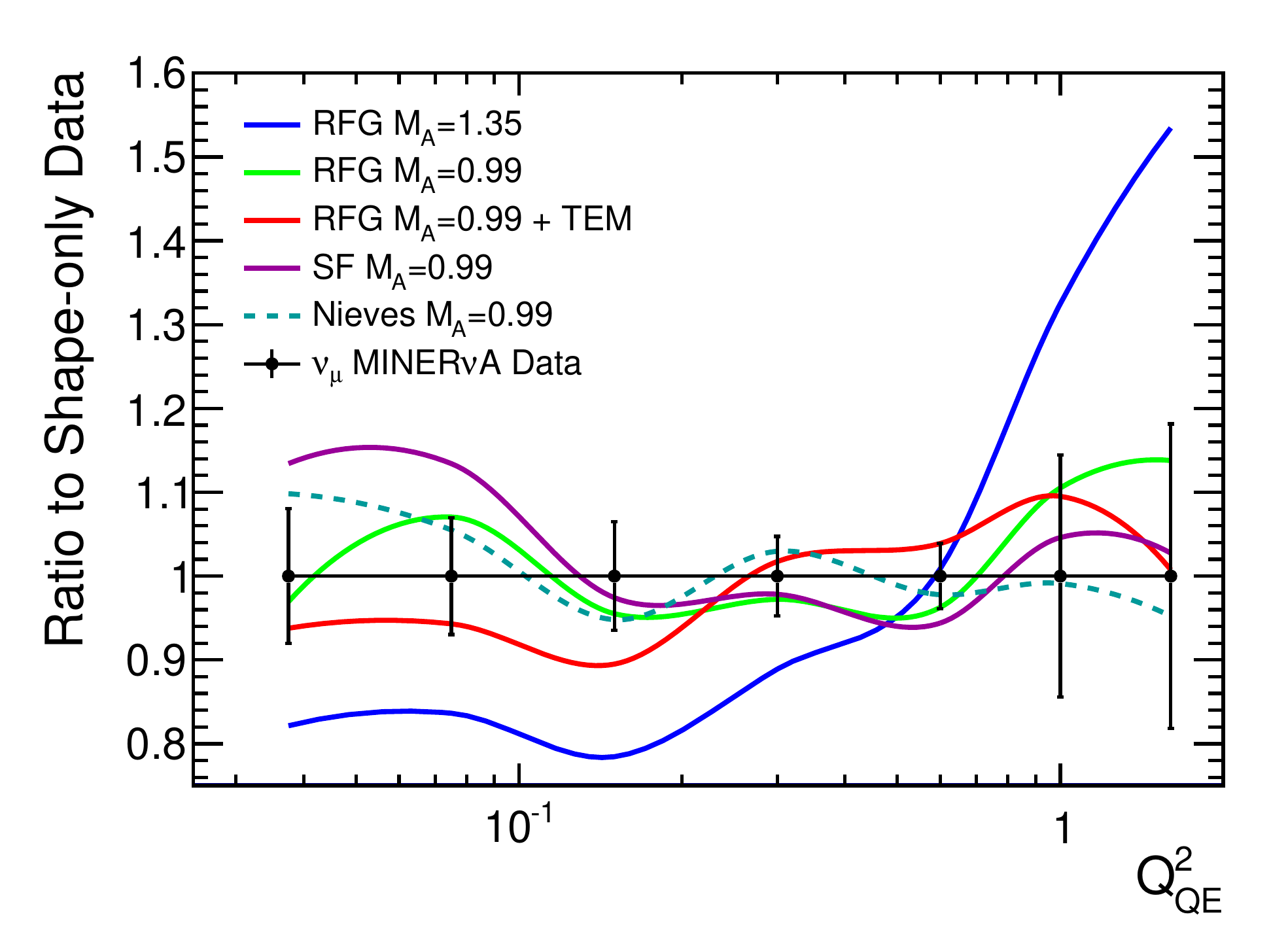} 
	\hfill
	\includegraphics[width=.49\textwidth, trim = 18 3 5 15, clip]{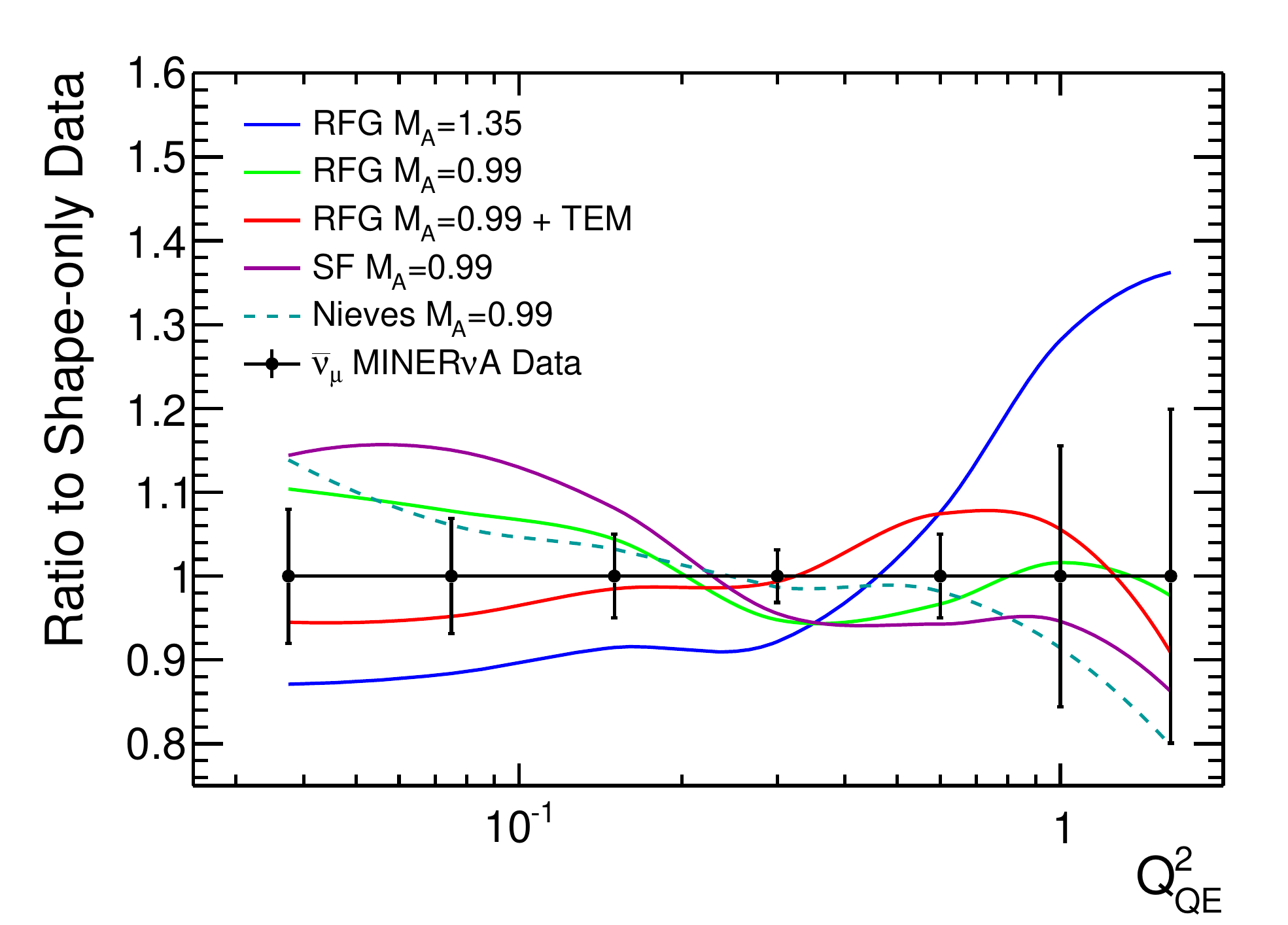} 

	\caption{Neutrino (left) and antineutrino (right) flux-averaged cross-section predictions are shown for all cross-section models investigated without
		    any sterile neutrino bias. The upper panel shows the differential cross-section and the lower panel the ratio of model to data. In the ratio model and data are area normalised.
		    The error bars on the MINER$\nu$A data include both statistical and systematic errors.} 
	\label{fig:shapecurves} 
\end{figure}

\begin{figure}[tbp] 
	\centering
	\includegraphics[width=.49\textwidth, trim = 10 3 20 10,clip]{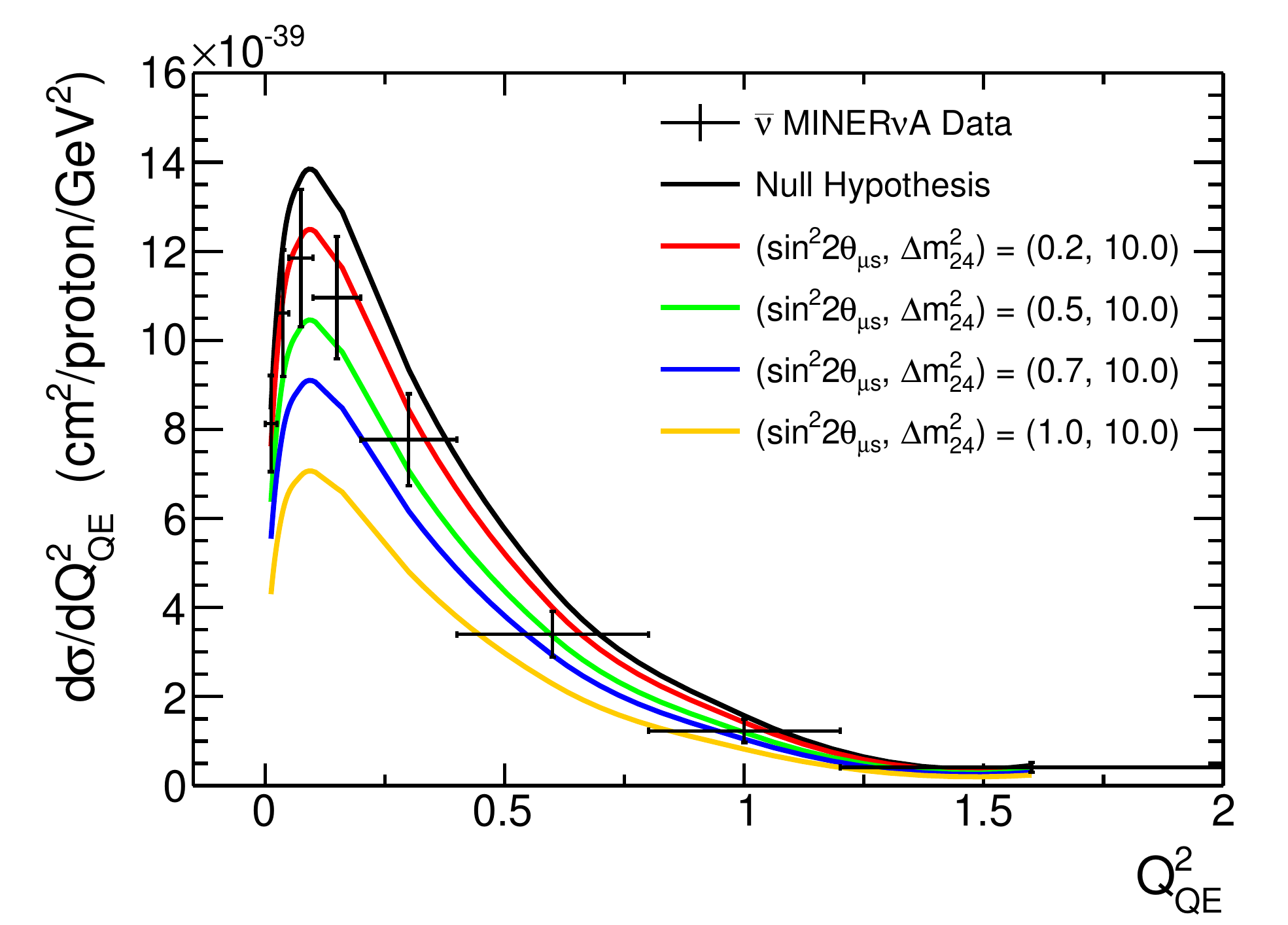} 
	\hfill 
	\includegraphics[width=.49\textwidth, trim = 10 3 20 10, clip]{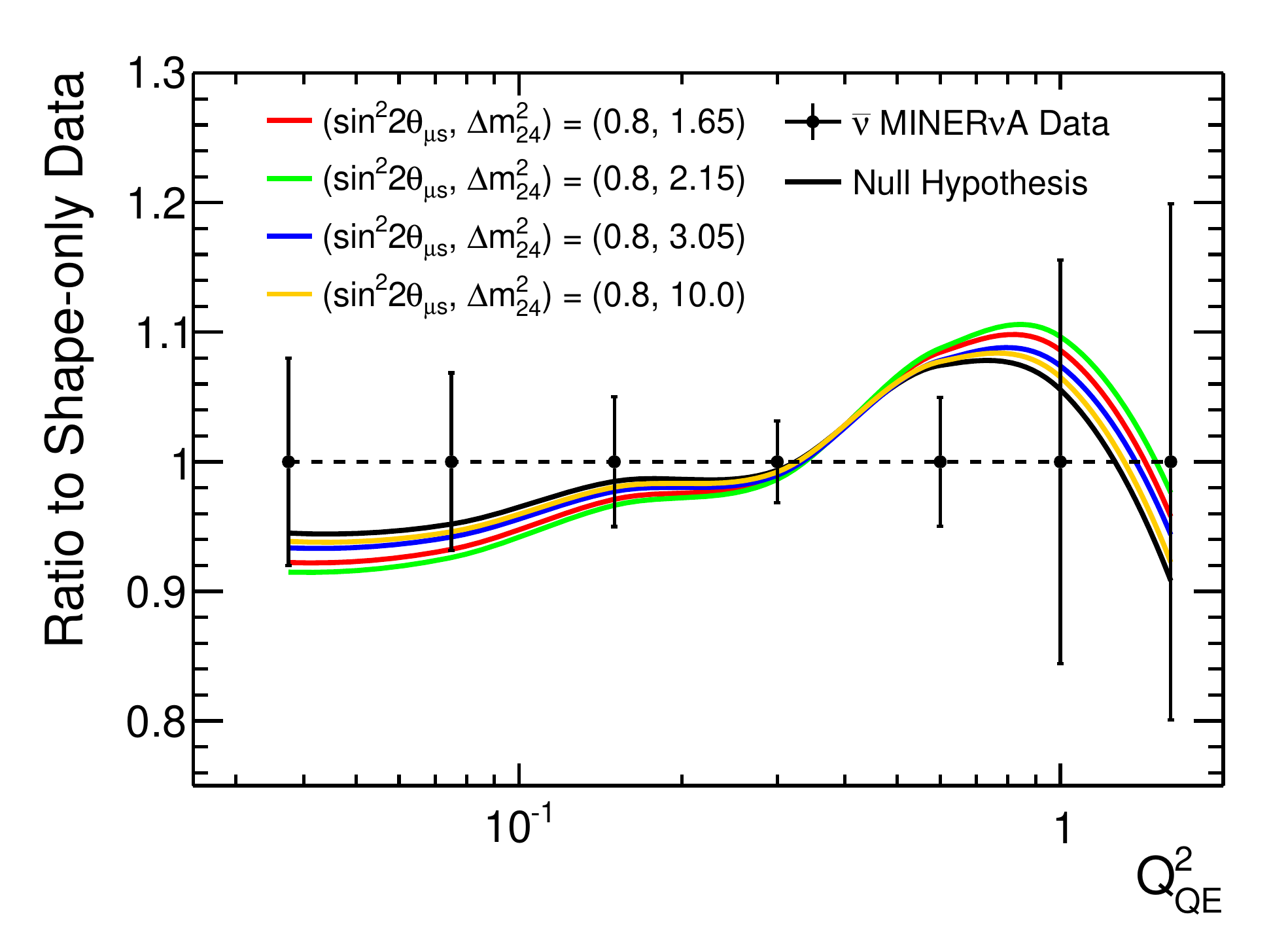} 
	\caption{Example sterile hypotheses for the RFG+TEM antineutrino model. Ratio to shape-only data (area normalised to unity)
	 is shown to highlight the limited shape sensitivity to sterile modifications.}
	\label{fig:nomrate}
\end{figure}

Unbiased cross-section predictions corresponding to the null hypothesis are shown for each
cross-section model in Figure \ref{fig:shapecurves}. The effect of sterile neutrino induced biases
on the RFG + TEM model over a range of mixing parameters can be seen in Figure \ref{fig:nomrate}.
Changes in $E_\nu$ distributions introduced by the bias have only a small effect on the shape
because the peak neutrino energy is higher than the experimental kinematic limit $2 m_p E_\nu > Q^2$
\cite{mcfarland11}. The shape's response to sterile modifications is likely to be larger for other
experiments.

The initial $\chi^2$ definition used to fit the sterile hypotheses to  data is 
\begin{align}
	\label{eq:chi2orig}
	\chi^{2} = \sum_{i=1}^{16} \sum_{j=1}^{16} 
	\left(\nu_i^{M} - \nu_i^{D} \right) M_{ij}^{-1}
	\left(\nu_j^{M} - \nu_j^{D}\right) \,,
\end{align} 
where $M_{ij}$ is the covariance matrix, $\nu_i^{D}$ are the measured differential
cross-section in $Q^2_{QE}$ bins, and $\nu_j^{M}$ are calculated using Equation
\eqref{eq:bincontents}. It was found that minimizing this statistic gave results far below the data
points, an effect consistent with ``Peelle's Pertinent Puzzle'' (PPP) \cite{peelle87puzzle}.

PPP can occur when fitting to a dataset containing large correlated uncertainties between all bins.
If the total normalisation is reduced in the fit, the relative size of the shape errors increases,
thus appearing to improve the agreement even if the shape of the prediction has not changed
\cite{peelle87puzzle}. This causes the fit to prefer parameter values which predict a distribution
that lies far below the data \cite{carlson09peelle}. We avoid the PPP problem by separating the
MINER$\nu$A covariance matrix into a total normalisation error, $\epsilon = 10.9\%$, and a
shape-only matrix, $M_{ij}^{shape}$ \cite{agostini94covar}.

The extracted shape-only covariance matrix, $M_{ij}^{shape}$, could not be inverted analytically as
a result of rounding errors in the published MINER$\nu$A data. We dealt with this problem by using
the two-step method of ref \cite{cteq01parton}, in which the bin errors and correlations are treated
separately. The alternative $\chi^2$ definition is given by
\begin{align} 
	\label{eq:eqchi2}
	\chi^{2} = 
	\left \lbrack \sum_{i=1}^{16} \left( \frac{ \nu_i^{D} - (\nu_i^{M} / \alpha)}
	{\sigma_{i}} \right)^2 - \sum_{i=1}^{16}\sum_{j=1}^{16} C_i(A^{-1})_{ij}C_j \right \rbrack 
	+ \left(\frac{1-\alpha}{\epsilon}\right)^2 \,,
\end{align} 
where $\sigma_i$ are the uncorrelated shape-only statistical errors for the dataset. The shape-only
correlated uncertainties are contained in $ C_i(A^{-1})_{ij}C_j$ which is defined in Appendix
\ref{app:fakedata}. The advantage of this procedure is that $A_{ij}$ is an invertible matrix. The
constrained parameter $\alpha$ normalised the theoretical predictions to the total measured
cross-section in the experimental range, allowing the square bracket to represent shape-only
contributions while the final penalty term reflected the difference in normalisation between the MC
and data. The combination of these techniques was found to be a robust way to fit a highly
correlated dataset affected by PPP (for a more detailed explanation of PPP and the definition of
\eqref{eq:eqchi2}, see Appendix \ref{app:fakedata}).

\begin{figure}[tbp] 
	\centering 
	\includegraphics[width=.495\textwidth,trim=50 20 30 5,clip]{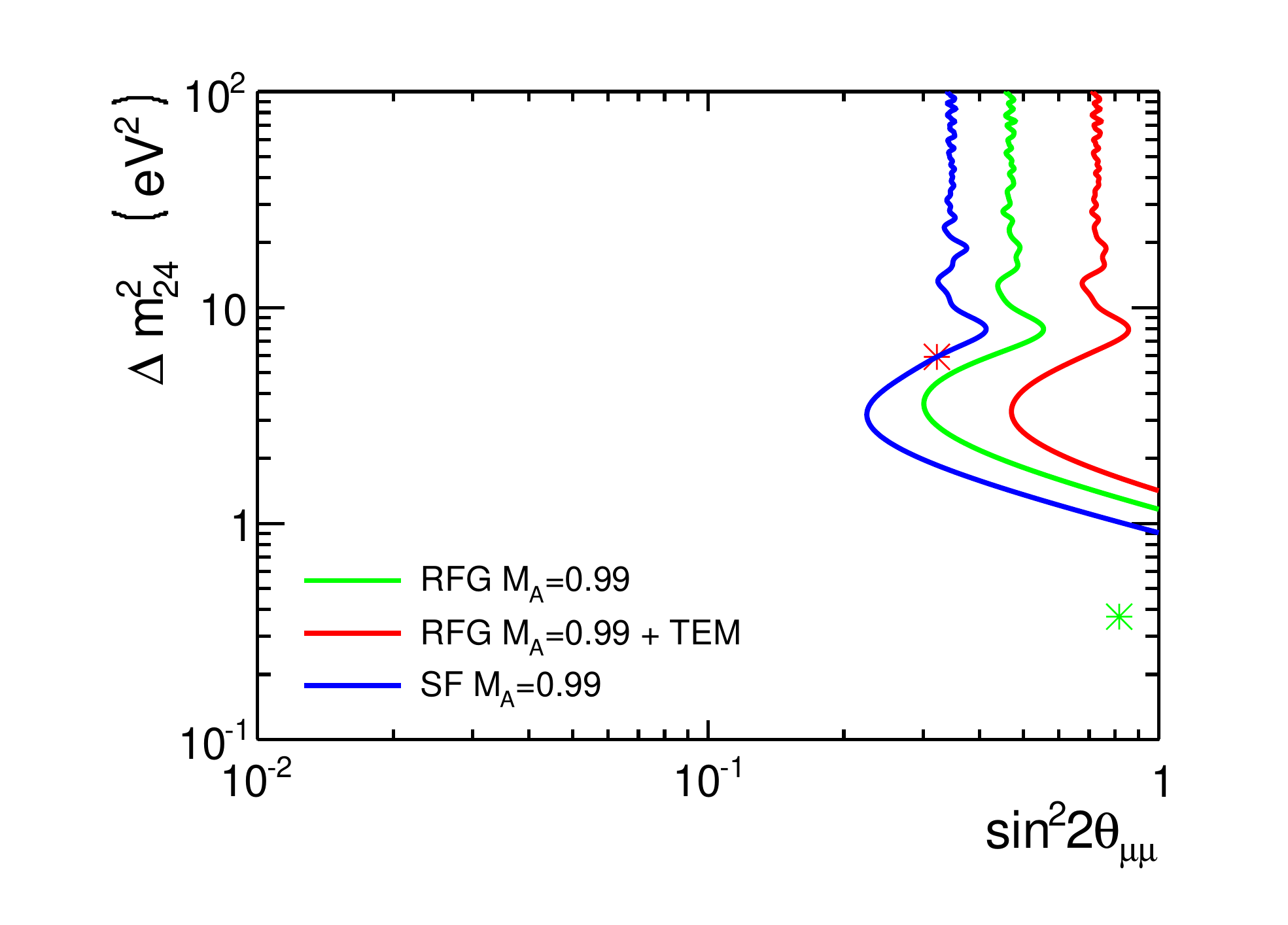}
	\hfill 
	\includegraphics[width=.495\textwidth,trim=50 20 30 5,clip]{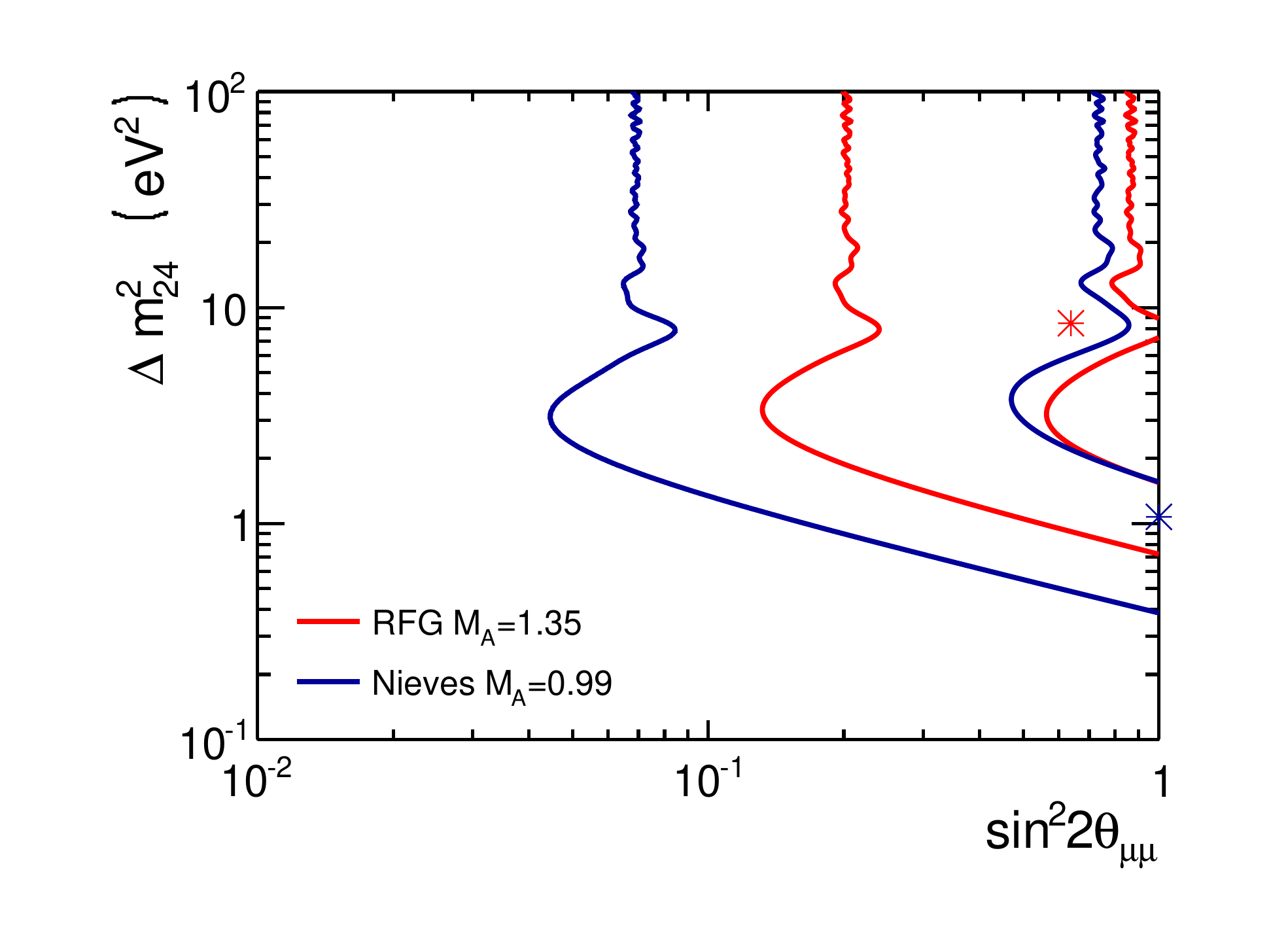} 
	\caption{90\% CL mixing parameter contours and best fit points (starred) for the cross-section models investigated.} 
	\label{fig:contours} 
\end{figure}

\begin{figure}[tbp] 
	\centering 
	\includegraphics[width=.495\textwidth,trim=40 20 20 30,clip]{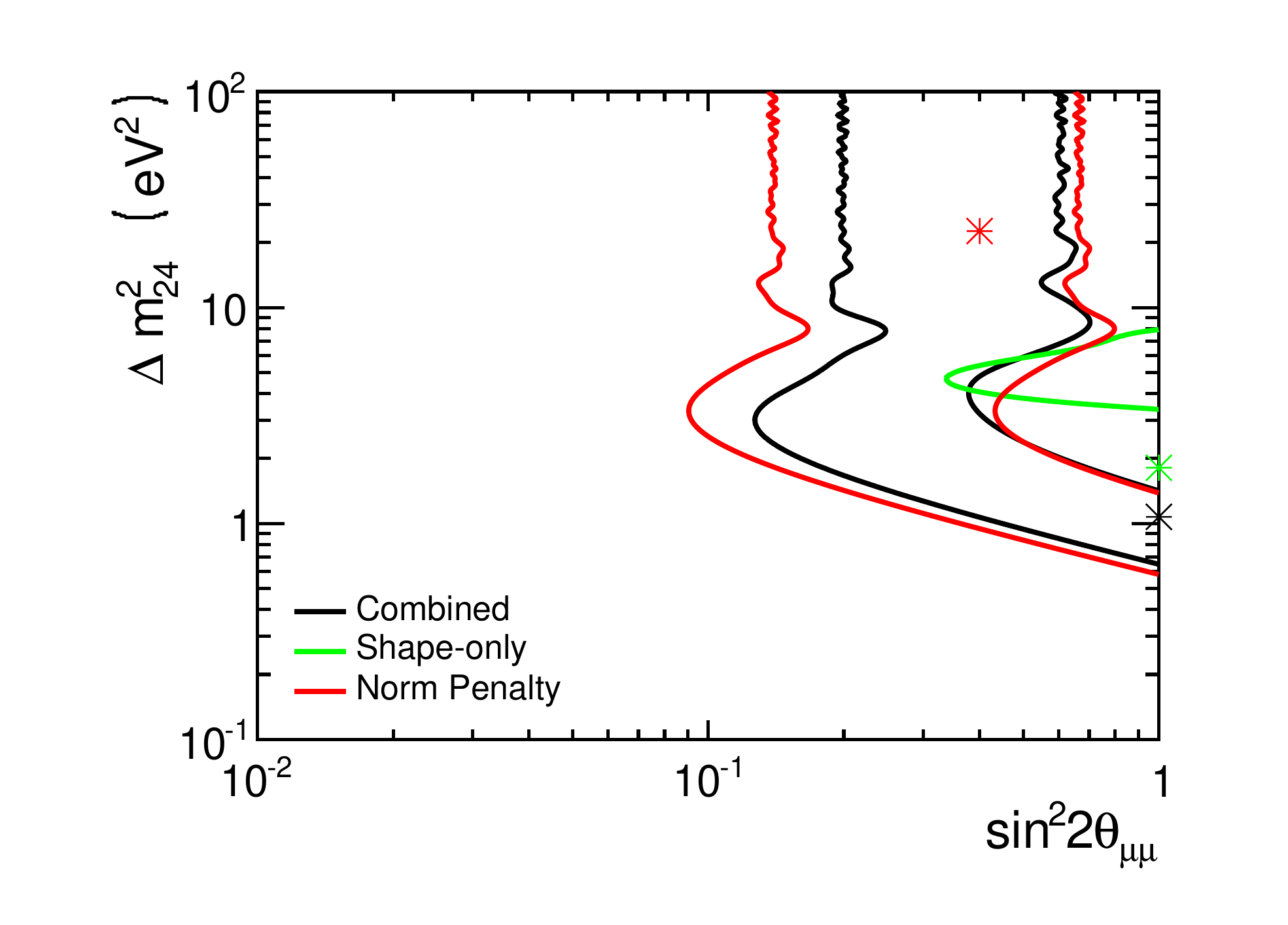} 
	\hfill 
	\includegraphics[width=.495\textwidth,trim=40 20 20 30,clip]{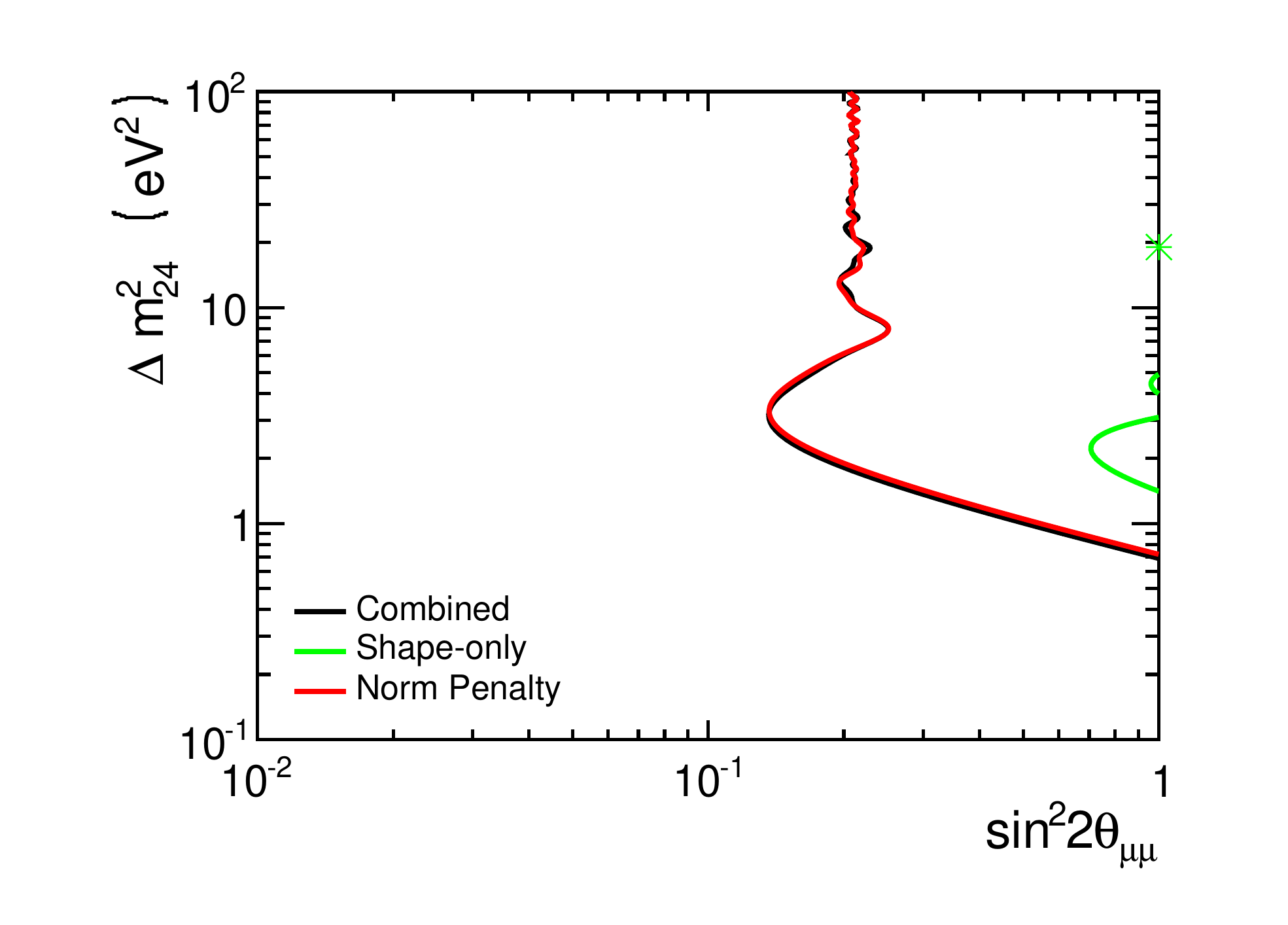}
	\\ 
	\caption{Overlaid shape-only, normalisation-only, and combined contours at the $1\sigma$ confidence limit. 
		    The Nieves (left) and SF (right) models are shown. Best fit points are indicated by a star.} 
	\label{fig:overlay} 
\end{figure}

For each cross-section model, parameter scans were performed for values in the ranges 0.0 $\leq
\sin^{2}2\theta_{\mu \mu} \leq$ 1.0 over 500 evenly spaced bins and $\mbox{0.1 eV}^2/\mbox{c}^4 \leq
\Delta m_{24}^{2} \leq \mbox{100 eV}^2/\mbox{c}^4$ over 1000 logarithmic bins. For each parameter
bin a sterile bias was introduced using the bin centre co-ordinates and a $\chi^2$ value calculated
using Equation \eqref{eq:eqchi2}. The minimum $\chi^2$ values from the scans were passed as starting
assumptions to MINUIT which then found a true $\chi^2$ minimum in the parameter space
\cite{minuit75,brun97root}. Best fit points and minimum $\chi^2$ values for each model can be found
in Table \ref{tab:chi2}. A $\Delta \chi^2$ method was used to produce 90\% CL confidence limits
around the best fit points as shown for all cross-section models in Figure \ref{fig:contours}. The
$1\sigma$ confidence limits for the separate shape-only or normalisation penalty terms from Equation
\eqref{eq:eqchi2} are compared in Figure \ref{fig:overlay} for the Nieves and SF models to highlight
the relative strength of the normalisation term.

\begin{table}[tbp] 
	\centering
	\begin{tabular}{ | l | c c c c c | }
		\hline
		\bf Model & \bf RFG 1.35 & \bf RFG 0.99 & \bf TEM 0.99 & \bf SF 0.99 & \bf NEV 0.99 \\
		\hline
		Nucleon distribution & RFG & RFG & RFG & SF & LFG \\
		$M_\mathrm{A}$ (GeV/c$^2$) & 1.35 & 0.99 & 0.99 & 0.99 & 0.99 \\
		Enhancements & - & - & TEM & - & Nieves + RPA \\
		\hline
		Null $\chi^2/15$ & 2.332 & 2.433 & 1.663 & 2.833 & 2.971 \\ 
		\hline
		Best sterile $\chi^2/13$ & 1.803 & 2.803 & 1.628 & 3.253 & 2.943 \\
		Best $\sin^2 2\theta_{\mu \mu}$ & 0.638 & 0.817 & 0.322 & 0.000 & 1.000 \\ 
		Best $\Delta m^{2}_{24}$ (eV$^2$)& 8.463 & 0.370 & 5.913 & 0.104 & 1.073 \\ 
		\hline 
	\end{tabular}
	\caption{Null hypothesis and best fit sterile hypothesis $\chi^2/Ndof$ values. }
	\label{tab:chi2}
\end{table}

\section{Discussion and conclusion} 
\label{sec:conclusion}

The $\chi^2$ values in Table \ref{tab:chi2} and the contours in figures \ref{fig:contours} and
\ref{fig:overlay} demonstrate the sensitivity of sterile neutrino fits to the adopted cross-section
model. Some models, e.g.\ RFG 1.35, exclude the null hypothesis at >99\%CL, while others, e.g.\ SF
0.99, prefer it. The choice of axial mass value is a critical parameter (compare RFG 1.35 with RFG
0.99), but other features of the models tested also have significant effects (consider RFG 0.99 and
TEM 0.99). It is clear that in many situations the choice of cross-section model can completely
dominate the results obtained in sterile neutrino searches.

In the case of the MINER$\nu$A dataset considered in this study, the weakness of the final limits
can be attributed to the large normalisation error. It is worth noting that the magnitude of the
disagreement between models is likely to increase when analysing data with a smaller normalisation
error or a stronger shape response to sterile neutrino biases. We conclude that sterile mixing
limits obtained in this way are subject to large systematic uncertainties, until they can be
repeated with a well-motivated theoretical model that agrees well with existing neutrino data.

\appendix

\section{$\chi^2$ definition and fake data study} 
\label{app:fakedata}

There is a well documented problem that can arise when fitting data with a covariance matrix that
contains large correlated uncertainties between bins, as in Equation \eqref{eq:appchi2orig}
\cite{agostini94covar}. By suppressing the normalisation of the prediction the $\chi^2$ is reduced,
leading to a best fit distribution well below the data. This occurs because the covariance matrix is
evaluated at a single point and the shape-only errors do not scale with normalisation. This problem
is known as ``Peelle's Pertinent Puzzle'' (PPP) \cite{peelle87puzzle}.
\begin{align}
	\label{eq:appchi2orig} \chi^{2} = \sum_{i=1}^{16} \sum_{j=1}^{16} 
	\left(\nu_i^{M} - \nu_i^{D} \right) M_{ij}^{-1}
	\left(\nu_j^{M} - \nu_j^{D}\right)
\end{align} 
Protecting against PPP is particularly important for sterile neutrino analyses, where a signal would
involve a suppression of the overall normalisation, and a large normalisation uncertainty due to
uncertainties in the flux prediction is common for accelerator experiments. PPP can be overcome by
redefining the $\chi^2$ in terms of the shape-only matrix, $M_{ij}^{shape}$, and scaling the total
integrated MC cross-section to match the total integrated cross-section in the data. This definition
effectively stops the fit from inflating the relative size of the shape-only errors
\cite{agostini94covar}.

\begin{align} 
\label{eq:appchi2shape}
	 \chi^{2} = \sum_{i=1}^{16} \sum_{j=1}^{16} 
	\left(\nu_i^{M} - \nu_i^{D} \right) \left(M_{shape}^{-1}\right)_{ij}
	\left(\nu_j^{M} - \nu_j^{D}\right)
\end{align} 

We obtain $M_{ij}^{shape}$ and the total normalisation uncertainty $\epsilon$ from the published
matrix $M_{ij}$ using the MiniBooNE matrix seperation method reproduced in Equation
\eqref{eq:miniboone} \cite{teppei08extract}.

\begin{align}
\nu_T &= \sum_k^{16} \nu_k^{D} \,, \quad \quad \alpha = \frac{\sum_i^{16} \nu_i^{M}}{\nu_T} \,, \notag \\
\epsilon &= \frac{1}{\nu_T} \sqrt{M_{ij}^{norm}} = \frac{1}{\nu_T} \sqrt{\sum_{k}^{16}\sum_{l}^{16} M_{kl}} \,,\notag\\
\label{eq:miniboone} M_{ij}^{shape} &=  M_{ij} - \frac{\nu^D_i}{\nu_T} \sum_k^{16} M_{kj} - \frac{\nu^D_j}{\nu_T} \sum_k^{16} M_{ik} 
	+ \frac{\nu^D_i \nu^D_j}{\nu_T^2} \sum_{k}^{16}\sum_{l}^{16} M_{kl} \,.
\end{align}

The matrix separation method involves summations over many matrix elements which can lead to large
rounding errors in $M_{ij}^{shape}$ if $M_{ij}$ is given to limited precision, as is the case for
the MINER$\nu$A data release. This can cause problems when inverting $M_{ij}^{shape}$. We modify the
$\chi^2$ definition in Equation \eqref{eq:appchi2shape} to avoid inverting the matrix
$M_{ij}^{shape}$ using the method given in ref. \cite{cteq01parton}. The final $\chi^2$ definition
used in our fits is given by
\begin{align} 
\label{eq:appchi2}
\chi^{2} = \left \lbrack \sum_{i=1}^{16} \left( \frac{ \nu_{i}^{D} - (\nu_{i}^{M} / \alpha)} {\sigma_{i}} \right)^2 
- \sum_{i=1}^{16}\sum_{j=1}^{16}C_i(A^{-1})_{ij}C_j \right \rbrack  
+  \left( \frac{1-\alpha}{\epsilon}\right)^2 \,,
\end{align} 
where $\sigma_i$ is the uncorrelated shape-only statistical error on the $i^{th}$ bin, and
$\Delta_{ik}$ is the correlated shape-only systematic uncertainty between the $i^{th}$ and $k^{th}$
bins which can be calculated using $\sum_{k=1}^{16} \Delta_{ik}\Delta_{kj} = M_{ij}^{shape} -
\sigma^2_{i} \delta_{ij}$. The vector $C$ and matrix $A$ are defined by
\begin{align} 
	\label{eq:ACeq}C_{i} =
	\sum_{k=1}^{16} \frac{\Delta_{ik} (\nu^D_i - (\nu^M_i/\alpha))}{\sigma_{k}^2} \,,
	\quad\quad
	A_{ij} =\delta_{ij} + \sum_{k=1}^{16} \frac{\Delta_{ik} \Delta_{jk}}{\sigma_{k}^2} \,.	
\end{align} 
The matrix $A$, which is inverted in this method, is less susceptible to the rounding error problems
than the full matrix $M_{ij}^{shape}$. Note that with more complete information the matrix $A$ can
be reduced to a $N \times N$ matrix for $N$ systematic errors as described in ref.
\cite{cteq01parton}.

To test the $\chi^2$ statistic as a method of fitting for sterile induced biases, we conducted a
fake data study using the RFG 1.35 dataset. We generated 30 sets of sterile parameters and generated
fake data for each using the following method.

First, MC was generated with the given parameter values according to the method described in Section
\ref{sec:method}. Then the systematic covariance matrix was calculated from the MINER$\nu$A matrix,
$M_{ij}$, and added to a diagonal matrix representing the statistical variance for the fake data
study to create a fake data covariance matrix, $M_{ij}^F$, as shown in Equation \eqref{eq:scalsys}.
This covariance matrix would have been produced if the fake data reflected nature, and was measured
in the MINER$\nu$A detector.
\begin{align} 
\label{eq:scalsys}M^F_{ij} &=  \frac{\nu^{M}_i \nu^{M}_j}{\nu^{D}_i \nu^{D}_j}
\left \lbrack M_{ij} - \delta_{ij}\left((\sigma^{D})_i^2 - (\sigma^{MC})_i^2\right) \right \rbrack
\end{align}
Throwing the matrix using the Cholesky decomposition method and adding the result to the nominal
fake data produces a realistic fake data sample \cite{efron1994introduction}. The residuals from
2000 throws were calculated and fitted with a Gaussian to look for biases for each of the 30 sterile
parameter sets investigated. These were found to have pulls away from the true parameter in the
range of $-0.05$ to $0.10$ for $\sin^22\theta_{\mu \mu}$ and $-0.24$ to $0.28$ for $\Delta
m^2_{24}$, and widths in the range of $0.81$ to $0.98$ for $\sin^22\theta_{\mu \mu}$ and $0.75$ to
$0.98$ for $\Delta m^2_{24}$. It was concluded that the $\chi^2$ statistic is a good estimator of
the central value and was appropriate for the analysis.

\acknowledgments
P.S. and C.W. acknowledge the UK STFC for support with PhD Studentships. S.C. acknowledges on going support from the STFC.


\bibliographystyle{JHEP}

\bibliography{SterilesMinerva}{}

\end{document}